\documentclass[12pt]{article}
\usepackage{amsmath}
\usepackage{amsfonts}
\usepackage{amssymb}
\usepackage{graphicx}

\input{tcilatex}
\begin{document}

\begin{titlepage}
%\begin{flushright}
%2008/3/2
%\end{flushright}
\ \\
\begin{center}
\LARGE
{\bf
Quantum Energy Teleportation \\
in Spin Chain Systems
}
\end{center}
\ \\
\begin{center}
\large{
Masahiro Hotta
}\\
\ \\
\ \\
{\it
Department of Physics, Faculty of Science, Tohoku University,\\
Sendai, 980-8578, Japan\\
hotta@tuhep.phys.tohoku.ac.jp

}
\end{center}
\begin{abstract}
We propose a protocol of quantum energy teleportation that transports 
 energy in spin chains to distant sites by only local operations and classical communication. 
The protocol uses ground-state entanglement and localized negative-energy excitation,  
and the energy is teleported without breaking any physical laws including causality and local energy conservation.  
\end{abstract}
Keywords: entanglement, quantum teleportation, spin chain, quantum information theory, condensed matter physics
\end{titlepage}

\bigskip

\bigskip

\section{Introduction}

\ \newline

Quantum entanglement opens the door to several interesting quantum
phenomena, including quantum teleportation (QT) \cite{qt}, by which an
unknown quantum state can be teleported to a distant place by local
operations and classical communication (LOCC). However, QT cannot transport
the excitation energy of the transported states. For example, let us imagine
that Alice sends to Bob a spin-up state of a spin in an external uniform
magnetic field parallel to the z axis. The Hamiltonian is given by 
\begin{equation*}
H_{s}=g\sigma _{z}
\end{equation*}%
with a positive constant $g$. The ground state of the spin is the spin-down
state with energy $-g$ and the energy of the excited spin-up state is $+g$.
A contracted state at Alice or Bob of a Bell pair shared by Alice and Bob
for QT is the maximal entropy state, which has zero energy on average.
Therefore, both Alice and Bob must first supply energy $+g$ to their spins
on average to create the Bell pair from two spins in the ground state by a
global operation. Bob has to supply an additional $+g$ energy to his spin to
receive the spin-up state from Alice's site by QT. Hence, the total energy
input at Bob's site is given by $+2g$, which should be locally prepared by
Bob.

Does the above-described protocol of QT imply that energy cannot be
transported by LOCC? \ Amazingly, the answer is ``no''. \ Quantum mechanics
allows energy transport by LOCC. In this paper, we propose a protocol of
quantum energy teleportation (QET) in spin chains which transports energy
only by LOCC, using local excitations with negative energy and ground-state
entanglement, thereby respecting fundamental physical laws including
causality and local energy conservation.

Spin chains are composed of many spins arrayed in one dimension. Short-range
interactions exist between the spins, and the Hamiltonian is given by a sum
of these local interaction terms. Due to the interactions, complicated
entanglement among the spins can occur even in the ground state.\ Spin
chains have recently attracted much attention in the context of quantum
information theory because they can be applied to the short transmission of
quantum states \cite{stqi}. Spin-chain entanglement is also helpful for
investigating the complicated physical properties of the ground state of
spin chains \cite{rev}.

Negative localized energy plays an essential role in the protocol. In this
paper, we define the zero values of energy density by the expectation values
of the ground state. We then consider why there are negative-energy-density
regions in the spin chains. In quantum physics, there remain local quantum
fluctuations even in the ground states of spin chains. By linearly
superposing the eigenstates of the total Hamiltonian, we can suppress the
quantum fluctuations more strongly in a local region via quantum
interference, as compared with the ground state. Taking a definition such
that the energy density in the ground state is zero, the energy density
becomes negative in a region where the quantum fluctuation is more greatly
suppressed. We might be concerned that there are states with energy values
lower than the ground state. However, it should be stressed that, even
though some regions have negative energy density, the total energy of the
system is always non-negative. Hence, there exist no states with energy
lower than the ground states. Negative energy effects in relativistic field
theory have long been investigated \cite{BD}. However, they have not often
been applied to condensed matter physics, quantum optics or quantum
communication. As an exceptional example, a fundamental lower bound of
actuating energy for photon switching has recently been derived for input
waves with negative energy density by a gedanken experiment \cite{hotta}.

In this paper, we present a protocol for near-critical two-level spin chains
with nondegenerate ground states and large correlation lengths. In the
analysis we assume that the number of spins is quite large, but finite.
However, the results obtained in this paper can also be applied to the case
of infinite degrees of freedom as long as an infinite limit of the number of
spins is justified. We also concentrate on short time scales in which
dynamical evolution induced by the Hamiltonian $H$ is negligible. We denote
the difference between the largest eigenvalue and the smallest eigenvalue of 
$H$ by $\Delta E$. The timescale $t$ considered is assumed to satisfy

\begin{equation}
t\ll \frac{1}{\Delta E}.  \label{sts}
\end{equation}%
Assuming this condition, it is valid to treat the time evolution operator as 
$\exp \left[ -itH\right] $ $\sim I$. It should also be noted that the
condition in eq. (\ref{sts}) can be weakened if a finite amount of energy $%
E_{in}$ less than $\Delta E$ is input to the spin chain by the energy sender
as follows:%
\begin{equation}
t\ll \frac{1}{E_{in}}.  \label{sts2}
\end{equation}%
On the other hand, we also assume that LOCC for the spins can be repeated
many times even in a short time interval. \ Taking the site number
difference between the two parties in the protocol as $\Delta n$ and the
lattice spacing between nearest-neighbor sites as $a$, the time scale
condition for many-round LOCC is expressed as

\begin{equation}
t\gg a\Delta n/c,  \label{locc}
\end{equation}%
where $c$ is the light velocity. By taking the nonrelativistic limit $%
c\rightarrow \infty $, the relation in eq. (\ref{locc}) always holds. The
proposed protocol is presented for one-dimensional qubit chain systems.
Extending the protocol to spin chain systems with larger spins and larger
dimensions is a straightforward task.

\bigskip

The paper is organized as follows. In section 2, we show that negative
localized energy density naturally appears in the systems under study. In
section 3, local energy conservation of the spin chain systems is discussed.
In section 4, we discuss the measurement of local observables for the ground
state, which is an important consideration for the QET protocol. In section
5, we propose the QET protocol. We conclude the paper in section 6.

\bigskip

\bigskip

\bigskip

\bigskip

\section{Ground-State Entanglement and Negative Energy Density\ \ }

\ \newline

Let us consider a spin chain with a nondegenerate ground state. The
Hamiltonian is given by a sum of semi-local components $T_{n}$: 
\begin{equation}
H=\sum_{n}T_{n}  \label{1}
\end{equation}%
Here $T_{n}$ are Hermitian operators given by

\begin{equation}
T_{n}=\sum_{\gamma }\prod_{m=n-L}^{n+L}O_{m}^{(n,\gamma )},  \label{1.1}
\end{equation}%
where $O_{m}^{(n,\gamma )}$ is a local Hermitian operator at site $m$ and
the integer $L$ denotes the interaction range. If we take $L=1$, the nearest
neighbor interaction can be treated. For example, the interaction of the
Ising model with a transverse magnetic field satisfies $L=1$ and has $T_{n}$
such that 
\begin{equation}
T_{n}=-b\sigma _{n}^{z}-\frac{h}{2}\sigma _{n}^{x}\left( \sigma
_{n+1}^{x}+\sigma _{n-1}^{x}\right) -\epsilon ,  \label{ising}
\end{equation}%
where $b$, $h$ and $\epsilon $ are real constants. $T_{n}$ of the example
has three terms in the right-hand side in eq. (\ref{1.1}) with $\gamma =1,2,3
$. These operators are given by

\begin{eqnarray*}
O_{n-1}^{(n,1)} &=&I, \\
O_{n}^{(n,1)} &=&-b\sigma _{n}^{z}-\epsilon , \\
O_{n+1}^{(n,1)} &=&I, \\
O_{n-1}^{(n,2)} &=&I, \\
O_{n}^{(n,2)} &=&-\frac{h}{2}\sigma _{n}^{x}, \\
O_{n+1}^{(n,2)} &=&\sigma _{n+1}^{x}, \\
O_{n-1}^{(n,3)} &=&\sigma _{n-1}^{x}, \\
O_{n}^{(n,3)} &=&-\frac{h}{2}\sigma _{n}^{x}, \\
O_{n+1}^{(n,3)} &=&I.
\end{eqnarray*}%
$T_{n}$ describes the local energy density at site $n$. The ground state $%
|g\rangle $ is an eigenstate for the lowest eigenvalue $E_{0}~$of $H$.
However, it is not guaranteed in general that $|g\rangle $ is an eigenstate
of $T_{n}$. Without changing the dynamics at all, it is always possible to
shift $T_{n}$ by adding constants. Hence, we are able to redefine $T_{n}$ to
satisfy 
\begin{equation}
\langle g|T_{n}|g\rangle =0.  \label{2}
\end{equation}%
Then, the shifted value of $E_{0}$ becomes zero as follows: 
\begin{equation*}
E_{0}=\langle g|H|g\rangle =\sum_{n}\langle g|T_{n}|g\rangle =0.
\end{equation*}%
From the above, it is satisfied that 
\begin{equation}
H|g\rangle =0  \label{3}
\end{equation}%
without any loss of generality. By this redefinition, $H$ becomes a
non-negative operator and satisfies 
\begin{equation}
\limfunc{Tr}\left[ \rho H\right] \geq 0  \label{4}
\end{equation}%
for an arbitrary quantum state $\rho $. This choice of the energy origin
simplifies the spin chain analysis, because we will often consider the
energy difference between an excited state and the ground state.

If each $T_{n}$ is an exact local operator at site $n$, all $T_{n}$ can be
simultaneously diagonalized and the ground state $|g\rangle $ is separable
and an eigenstate for the lowest eigenvalue of each $T_{n}$. In such a
situation, $T_{n}$ becomes non-negative. However, the condition is not
sustained in general for cases with interactions between spins, and
entangled ground states often appear. In the later discussion, we are
interested in spin-chain models with entangled ground states. Let us
consider an ordinary case of a correlation function $\langle
g|T_{n}O_{m}|g\rangle $, not decomposed into $\langle g|T_{n}|g\rangle
\langle g|O_{m}|g\rangle $:

\begin{equation}
\langle g|T_{n}O_{m}|g\rangle \neq \langle g|T_{n}|g\rangle \langle
g|O_{m}|g\rangle   \label{crf}
\end{equation}%
for a certain site $n$ and local operator $O_{m}$ at site $m$ with $%
\left\vert n-m\right\vert \geq L+1$. Clearly, the state $|g\rangle $ is
entangled because all separable ground states satisfy $\langle
g|T_{n}O_{m}|g\rangle =\langle g|T_{n}|g\rangle \langle g|O_{m}|g\rangle $.
It is proven easily from eq. (\ref{crf}) that the state $|g\rangle $ is not
an eigenstate of $T_{n}$. The reason is as follows. If $T_{n}|g\rangle
=c_{n}|g\rangle $ is satisfied for a certain real constant $c_{n}$, the
correlation function must be written as 
\begin{equation*}
\langle g|T_{n}O_{m}|g\rangle =c_{n}\langle g|O_{m}|g\rangle =\langle
g|T_{n}|g\rangle \langle g|O_{m}|g\rangle ,
\end{equation*}%
which contradicts eq. (\ref{crf}). Therefore we obtain the relation:

\begin{equation}
T_{n}|g\rangle \neq c_{n}|g\rangle   \label{nes}
\end{equation}%
for arbitrary real constant $c_{n}$. eq. (\ref{nes}) gives important
information about the emergence of a negative energy density as follows.
Because the operator $T_{n}$ is a Hermitian operator acting on the total
Hilbert space of the spin chain, $T_{n}$ can be spectrally decomposed into

\begin{equation*}
T_{n}=\sum_{\nu ,k_{\nu }}\epsilon _{\nu }(n)|\epsilon _{\nu }(n),k_{\nu
},n\rangle \langle \epsilon _{\nu }(n),k_{\nu },n|,
\end{equation*}%
where $\epsilon _{\nu }(n)$ are eigenvalues of $T_{n}$, $|\epsilon _{\nu
}(n),k_{\nu },n\rangle $ are corresponding eigenstates and the index $k_{\nu
}$ denotes the degeneracy freedom of the eigenvalue $\epsilon _{\nu }(n)$.
Because $\left\{ |\epsilon _{\nu }(n),k_{\nu },n\rangle \right\} $ is a
complete orthogonal set of basis state vectors of the total Hilbert space,
the ground state in the total Hilbert space can be uniquely expanded as%
\begin{equation}
|g\rangle =\sum_{\nu ,k_{\nu }}g_{\nu ,k_{\nu }}(n)|\epsilon _{\nu
}(n),k_{\nu },n\rangle .  \label{exg}
\end{equation}%
Using this expansion, eq. (\ref{2}) gives%
\begin{equation}
\langle g|T_{n}|g\rangle =\sum_{\nu ,k_{\nu }}\epsilon _{\nu }(n)\left\vert
g_{\nu ,k_{\nu }}(n)\right\vert ^{2}=0.  \label{geq}
\end{equation}%
Clearly, eq. (\ref{geq}) has no solutions if the lowest eigenvalue $\epsilon
_{-}(n)$ of $T_{n}$ is positive. For the case of $\epsilon _{-}(n)=0$, eq. (%
\ref{geq}) has a solution with $g_{-,k_{-}}(n)\neq 0$. The other components $%
g_{\nu ,k_{\nu }}(n)$ vanish. However, this solution implies that $%
T_{n}|g\rangle =c_{n}|g\rangle $ with $c_{n}=\epsilon _{-}(n)$ because of
eq. (\ref{exg}) and apparently contradicts eq. (\ref{nes}). Therefore $%
\epsilon _{-}(n)$ does not take a zero value. Hence, it is concluded that $%
\epsilon _{-}(n)$ must be negative:%
\begin{equation*}
\epsilon _{-}(n)=-\left\vert \epsilon _{-}(n)\right\vert <0.
\end{equation*}%
The average energy density for the corresponding eigenstate $|\epsilon
_{-}(n),k_{-},n\rangle $ also becomes negative. It is thereby verified that
there exist quantum states with negative energy density for spin chains
satisfying eq. (\ref{crf}). It should be stressed again that even if a state
has negative energy density over a certain region, there exists positive
energy density at other regions and the total energy is not negative because
of the non-negativity of $H$.

\bigskip

\section{Local Energy Conservation}

\ \newline

In this section, the local energy conservation of the spin chain system $S$
is explained. First, a free system, which is not coupled with external
systems, is considered. The system evolves by the Hamiltonian in eq. (\ref{1}%
). A connected site region is introduced, $V$, given by $[n_{i},n_{f}]$%
\thinspace\ that satisfies $n_{f}-n_{i}\geq 2L-1$. The energy of $V$ is
defined by

\begin{equation}
H_{V}=\sum_{n=n_{i}}^{n_{f}}T_{n}.  \label{hv}
\end{equation}%
In a free system, the Heisenberg operator of a Schr\"{o}dinger operator $O$
is defined by $O^{(H)}(t)=e^{itH}Oe^{-itH}$. The Heisenberg operator of $%
H_{V}$ evolves by the following equation: 
\begin{equation}
\frac{d}{dt}H_{V}^{(H)}=i[H,~H_{V}^{(H)}].  \label{heq}
\end{equation}%
By substituting a relation given by

\begin{equation*}
H=\sum_{n\leq
n_{i}-2L-1}T_{n}^{(H)}+\sum_{n=n_{i}-2L}^{n_{i}-1}T_{n}^{(H)}+H_{V}^{(H)}+%
\sum_{n=n_{f}+1}^{n_{f}+2L}T_{n}^{(H)}+\sum_{n\geq n_{f}+2L+1}T_{n}^{(H)}
\end{equation*}%
into eq. (\ref{heq}), it is possible to obtain%
\begin{equation*}
\frac{d}{dt}H_{V}^{(H)}=i[\sum_{n=n_{i}-2L}^{n_{i}-1}T_{n}^{(H)}+%
\sum_{n=n_{f}+1}^{n_{f}+2L}T_{n}^{(H)},~H_{V}^{(H)}].
\end{equation*}%
Substituting eq. (\ref{hv}) into the above equation yields the following
energy conservation relation for an arbitrary initial state $|\psi \rangle $.%
\begin{equation*}
\frac{d}{dt}\langle \psi |H_{V}^{(H)}|\psi \rangle =J_{n_{i}-1}-J_{n_{f}},
\end{equation*}%
where the above energy fluxes are given by

\begin{equation}
J_{n}=i\langle \psi |[\sum_{m=n-2L+1}^{n}T_{m}^{(H)},~\sum_{m^{\prime
}=n+1}^{n+2L}T_{m^{\prime }}^{(H)}]|\psi \rangle  \label{j}
\end{equation}%
If $J_{n_{i}-1}=J_{n_{f}}=0$, the energy of $V$ does not change at all.

Next, a situation is considered in which the spin chain $S$ is locally
coupled with external systems $D$ and $C$. $S$ interacts with \ the external
systems only at site $n_{o}$. The coupling can be switched on and off
effectively by the time evolution of the switch system $C $. System $D$
locally controls the spin chain in the switch-on interval. The free part of
the total Hamiltonian is then written as

\begin{equation*}
H_{o}=H+H_{C}+H_{D},
\end{equation*}%
where $H_{C}$ is the free Hamiltonian of $C$, and $H_{D}$ is of $D$. In
order to capture the essence of physics, example cases are considered in
which the switch system $C$ is a one-dimensional bosonic Schr\"{o}dinger
field $\Psi (x)$ with Hamiltonian

\begin{eqnarray*}
H_{C} &=&-\frac{iv}{2}\int_{-\infty }^{\infty }\left[ \Psi ^{\dagger
}(x)\partial _{x}\Psi (x)-\partial _{x}\Psi ^{\dagger }(x)\Psi (x)\right] dx
\\
&&+\frac{1}{2M}\int_{-\infty }^{\infty }\partial _{x}\Psi ^{\dagger
}(x)\partial _{x}\Psi (x)dx
\end{eqnarray*}%
where$\ v$ and $M$ are positive real paramters and the operators $\Psi
(x)\,\ $and$~\Psi ^{\dag }(x)$ satisfy%
\begin{eqnarray*}
\left[ \Psi (x),~\Psi ^{\dag }(x^{\prime })\right] &=&i\delta \left(
x-x^{\prime }\right) , \\
\left[ \Psi (x),~\Psi (x^{\prime })\right] &=&0, \\
\left[ \Psi ^{\dag }(x),~\Psi ^{\dag }(x^{\prime })\right] &=&0.
\end{eqnarray*}%
In later discussion, we take the mass parameter $M$ very large. Then $v$
bocomes the propagating velocity of right-moving excitation of the field
outside the interaction region: 
\begin{equation*}
e^{itH_{o}}\Psi ^{\dag }(x)e^{-itH_{o}}\approx \Psi ^{\dag }(x-vt).
\end{equation*}
The vacuum state $|0\rangle $ of $\Psi $ is defined by $\Psi (x)|0\rangle
=0. $ By use of the explicit example of $C$, a model of local operation for
the spin $S$ at site $n_{o}$ with energy conservation can be made. The total
Hamiltonian is given by

\begin{eqnarray}
H_{tot} &=&H_{o}+H_{int}^{(n_{o})},  \notag \\
H_{int}^{(n_{o})} &=&gK_{S+D}^{(n_{o})}\int_{-d}^{d}\Psi ^{\dagger }(x)\Psi
(x)dx  \label{int}
\end{eqnarray}%
Here $g$ is a real coupling constant, $d$ fixes the interaction region of $%
\Psi $ to $S$ and $D$. $K_{S+D}^{(n_{o})}$ is a Hermitian operator acting on
the Hilbert space of the composite system $S$ and $D$, and can be decomposed
as

\begin{equation*}
K_{S+D}^{(n_{o})}=I_{n_{o}}O_{D}+\vec{\sigma}_{n_{o}}\cdot \vec{O}_{D},
\end{equation*}%
where $O_{D}$ and $\vec{O}_{D}$ are Hermitian operators acting on the
Hilbert space of $D$. In this model, energy conservation is satisfied via

\begin{equation}
\frac{d}{dt}\left( \hat{H}(t)+\hat{H}_{C}(t)+\hat{H}_{D}(t)+\hat{H}%
_{int}^{(n_{o})}(t)\right) =0,  \label{ec}
\end{equation}%
where the hat operators denote the Heisenberg operator corresponding to Schr%
\"{o}dinger operators:

\begin{equation*}
\hat{O}(t)=e^{itH_{tot}}Oe^{-itH_{tot}}.
\end{equation*}

Switching is realized by scattering a wave packet of $\Psi $ via the
interaction in eq. (\ref{int}). Let us set the initial state of $C$ to a
coherent state 
\begin{equation*}
|\phi _{C}\rangle =N_{\phi }\exp \left( \int_{x_{0}}^{x_{1}}\phi (x)\Psi
^{\dag }(x)dx\right) |0\rangle ,
\end{equation*}%
where $\phi (x)$ is amplitude of the coherent state and localized in the
space region $\left[ x_{0},x_{1}\right] $ such that $x_{1}<-d$. Let us
assume that the total initial state is a product state given by $|I\rangle
=|\psi _{S}\rangle |\psi _{D}\rangle |\phi _{C}\rangle ,$ where $|\psi
_{S}\rangle $ is the initial state of $S$ and $|\psi _{D}\rangle $ is the
initial state of $D$. It is noted that in the initial phase, the interaction
in eq. (\ref{int}) is switched off because no wave of $\Psi $ exists inside
the interaction region $\left[ -d,d\right] $ and $H_{int}^{(n_{o})}$ has no
contribution. The wave packet evolves freely with velocity $v$ until a part
of the packet reaches $\left[ -d,d\right] $. When a part of the wave packet
stays in $\left[ -d,d\right] $, the interaction in eq. (\ref{int}) is
effectively switched on and energy can be exchanged among $S,C$ and $D$. In
the interval, the wave packet of $\Psi $ is scattered by the interaction.
After the scattering process, the scattered wave of $\Psi $ escapes with
velocity $v$ from the interaction region $\left[ -d,d\right] $ and $%
H_{int}^{(n_{o})}$ again has no contribution. Hence, taking account of the
disappearance of $H_{int}^{(n_{o})}$ contribution \ in the past and future
of the scattering events, energy conservation in eq. (\ref{ec}) clearly
yields the following relation: 
\begin{equation}
\langle I|\left( \hat{H}(t_{i})+\hat{H}_{C}(t_{i})+\hat{H}_{D}(t_{i})\right)
|I\rangle =\langle I|\left( \hat{H}(t_{f})+\hat{H}_{C}(t_{f})+\hat{H}%
_{D}(t_{f})\right) |I\rangle ,  \label{ec1}
\end{equation}%
where $t_{i}$ is the initial time, and $t_{f}$ is the final time of the
scattering process. It should be stressed that the scattering interval $%
t_{f}-t_{i}~$can go to zero as $v$ and $g$ approach infinity. It should be
recalled that the nonrelativistic limit is assumed in this paper, and the
speed of light is assumed to be infinity. Thus, $v$ can also approach
infinity. At this limit, the time evolution of the spins of $S$ at all sites
except site $n_{o}$ can be neglected during the short-time scattering
process. Then, energy exchange occurs not globally but locally among spins
around site $n_{o}$, $C$ and $D$. Therefore, eq. (\ref{ec1}) can be
rewritten as 
\begin{eqnarray}
&&\langle I|\left( \sum_{n=n_{o}-L}^{n_{o}+L}\hat{T}_{n}(t_{i})+\hat{H}%
_{C}(t_{i})+\hat{H}_{D}(t_{i})\right) |I\rangle   \notag \\
&=&\langle I|\left( \sum_{n=n_{o}-L}^{n_{o}+L}\hat{T}_{n}(t_{f})+\hat{H}%
_{C}(t_{f})+\hat{H}_{D}(t_{f})\right) |I\rangle .  \label{ec2}
\end{eqnarray}%
If $\langle I|\left( \hat{H}(t_{f})-\hat{H}(t_{i})\right) |I\rangle $ is
positive, the spin chain gets energy from $C$ and $D$ in the short-time
scattering interval. Inversely, if $\langle I|\left( \hat{H}(t_{f})-\hat{H}%
(t_{i})\right) |I\rangle $ is negative, the spin chain gives energy to the
composite system of $C$ and $D$. This fact becomes significant in the QET
protocol proposed later.

\section{Local Measurement of the Ground State and Entanglement Breaking}

\bigskip

For the QET protocol, it is important to consider the local measurement of
an entangled ground state $|g\rangle $ of a spin chain. Let us consider a
spin at site $n_{A}$ in the spin chain and a Hermitian unitary local
operator $\sigma _{A}=\vec{u}_{A}\cdot \vec{\sigma}_{n_{A}}$. Here $\vec{u}%
_{A}$ is a three-dimensional real unit vector and $\vec{\sigma}_{n_{A}}$ is
the Pauli spin matrix vector at site $n_{A}$. The eigenvalues of $\sigma
_{A} $ are $\left( -1\right) ^{\mu }$ with $\mu =0,1$. The spectral
expansion of $\sigma _{A}$ is given by 
\begin{equation}
\sigma _{A}=\vec{u}_{A}\cdot \vec{\sigma}_{A}=\sum_{\mu =0,1}\left(
-1\right) ^{\mu }P_{A}\left( \mu \right) ,  \label{uas}
\end{equation}%
where $P_{A}(\mu )$ is a projective operator onto the eigensubspace with $%
\mu $. Consider that Alice at site $n_{A}$ performs a projective measurement
of $\sigma _{A}$ for the ground state $|g\rangle $ and assume that she
obtains the measurement result $\mu $. The post-measurement state with $\mu $
is given by%
\begin{equation}
\frac{1}{\sqrt{p_{A}(\mu )}}P_{A}\left( \mu \right) |g\rangle ,  \label{sa}
\end{equation}%
where $p_{A}(\mu )=\langle g|P_{A}\left( \mu \right) |g\rangle $. Therefore,
the average post-measurement state is a mixed quantum state $\rho ^{\prime }$
given by

\begin{eqnarray*}
\rho ^{\prime } &=&\sum_{\mu =0,1}p_{A}(\mu )\frac{1}{\sqrt{p_{A}(\mu )}}%
P_{A}\left( \mu \right) |g\rangle \langle g|P_{A}\left( \mu \right) \frac{1}{%
\sqrt{p_{A}(\mu )}} \\
&=&\sum_{\mu =0,1}P_{A}\left( \mu \right) |g\rangle \langle g|P_{A}\left(
\mu \right) .
\end{eqnarray*}%
It should be noted that $\rho ^{\prime }$ is a quantum state which satisfies

\begin{equation}
\limfunc{Tr}_{n_{A}}\left[ \rho ^{\prime }\right] =\limfunc{Tr}_{n_{A}}\left[
|g\rangle \langle g|\right] ,  \label{qfb}
\end{equation}%
where $\limfunc{Tr}_{n_{A}}$ means a partial trace in terms of the measured
spin at site $n_{A}$. Therefore, the quantum fluctuation of $\rho ^{\prime }$
is the same as that of the ground state except at site $n_{A}$. It should
also be noted that the above projective measurement needs energy. Because
the ground state has zero energy, the input $E_{A}$ is calculated as follows:

\begin{equation}
E_{A}=\limfunc{Tr}\left[ \rho ^{\prime }H\right] -\langle g|H|g\rangle
=\sum_{\mu =0,1}\langle g|P_{A}\left( \mu \right) HP_{A}\left( \mu \right)
|g\rangle .  \label{ea}
\end{equation}%
Because of the non-negativity of $H$, $E_{A}$ is non-negative. In general, $%
P_{A}\left( \mu \right) |g\rangle $ is not proportional to $|g\rangle $ due
to entanglement and we obtain a positive value of $E_{A}$. Because the
measurement is performed locally, the excitation energy of the spin chain is
localized around site $n_{A}$ soon after the measurement. For convenience,
we introduce a localized energy operator around site $n$ by

\begin{equation}
H_{n}=\sum_{m=n-L}^{n+L}T_{m}.  \label{le}
\end{equation}%
$H_{n}$ includes all $T\,_{m}$'s for which expectation values change if
quantum operations act on site $n$. From eq. (\ref{le}), it can be shown
that 
\begin{equation*}
\limfunc{Tr}\left[ \rho ^{\prime }H_{n_{A}}\right] =E_{A}.
\end{equation*}%
Also, it is proven that

\begin{equation}
\limfunc{Tr}\left[ \rho ^{\prime }T_{n}\right] =0  \label{l2}
\end{equation}%
where $|n-n_{A}|>L$. eq. (\ref{l2}) implies localization of the energy
distribution around site $n_{A}$.

Next, consider that Alice at site $n_{A}$ attempts to completely withdraw
the input energy $E_{A}~$by local operations soon after the measurement. Her
available processes are expressed by arbitrary trace-preserving
completely-positive (TPCP) maps $\Gamma _{A}$ of the spin state at site $%
n_{A}$. In spite of any use of local TPCP maps, her attempt cannot be
achieved because the local measurement breaks the entanglement between the
spin at site $n_{A}$ and different-site spins in the spin chain. If Alice
wants to recover the original ground state, she must recreate the
entanglement broken by the first measurement. However, entanglement
generation needs nonlocal operations in general. Therefore, she cannot
recover the ground state perfectly by her local operations alone. This
observation implies that there is nonvanishing residual energy $E_{res}$ in
Alice's local cooling process to extract the input energy:

\begin{equation*}
E_{res}=\min_{\Gamma _{A}}\left( \limfunc{Tr}\left[ \Gamma _{A}\left[ \rho
^{\prime }\right] H_{n_{A}}\right] \right) >0.
\end{equation*}%
Therefore, interestingly, Alice is not able to use the residual energy even
though it is right in front of her. In the next section, we argue that a
part of the residual energy can be extracted by Bob, who is situated apart
from Alice and performs local operations dependent on the measurement result 
$\mu $ of $\sigma _{A}$ to a spin on the chain in front of him. This
protocol is QET.

\bigskip

\section{Quantum Energy Teleportation}

\ \newline

\bigskip

In this section, we discuss QET for spin chain systems. As mentioned in the
introduction, we concentrate on short time scales, in which dynamical
evolution induced by the Hamiltonian of the spin chain is negligible. On the
other hand, we also assume the nonrelativistic limit that LOCC for the spins
can be repeated many times even in a short time interval. Initially, Alice
and Bob share many near-critical spin chains in the ground state $|g\rangle
, $ which is entangled and has a large correlation length $l$. Alice is
situated at site $n_{A}$ and Bob at site $n_{B}$. Alice is a good distance
from Bob: $\left\vert n_{A}-n_{B}\right\vert \sim O(l)\gg 1$. Consider a
protocol consisting of the following steps. (I) Alice performs a measurement
of a local observable of the spin at site $n_{A}$ and obtains the
measurement result $\mu $. In the measurement process, the spin chain is
locally excited with some energy input. (II) Alice announces to Bob the
result $\mu $ by a classical channel at the speed of light or near the light
velocity. Because the time interval of the annoucement is very short, the
system does not evolve and no energy flow appears between Alice and Bob.
(III) Bob performs a local operation dependent on $\mu $ to the spin at site 
$n_{B}$. QET is defined by this protocol if the expectation value of $%
H_{n_{B}}$ is negative after the local operation of Bob. Because the
expectation value of $H_{n_{B}}$ is zero just before Bob's operation and
local energy conservation holds around Bob's region as mentioned in section
3, the negative value after Bob's operation implies that energy release
occurs around site $n_{B}$ to external systems during the operation
interval. (For the case in section 3, the energy moves from the spin chain $%
S $ to the external systems $C$ and $D$.) If Alice inputs no energy into the
spin chain or Bob does not use the information about $\mu $, Bob is not able
to extract energy from the spin chain at \ all. Therefore, it can be said
that a part of the energy input by Alice's measurement at site $n_{A}$ is
effectively transported to site $n_{B}$ by LOCC. If we select as the initial
state of the spin chain, not the ground state but an excited state which has
nonzero energy distribution around Bob, it is not unusual that Bob's local
operation can extract energy from the spin chain. However, in QET, the
ground state is actually selected as the initial state and the local energy
extraction is nontrivial.

\bigskip

Next we propose an explicit protocol of QET. The protocol with $L=1$ is
illustrated in Figure 1.

\bigskip

(I) Alice performs a projective measurement of the observable $\sigma _{A}$
in eq. (\ref{uas}) for the ground state $|g\rangle $ and obtains the
measurement result $\mu =0$ or $1$. Alice inputs energy $E_{A}$ in eq. (\ref%
{ea}) to the spin chain system in order to achieve the local measurement, as
seen in section 4.

\bigskip

(II)Alice announces to Bob the result $\mu $ by a classical channel. In the
announcement process, both time evolution of the system and emergence of
energy flux do not happen because the time interval is assumed very short.

\bigskip

(III) To a spin at site $n_{B}$, Bob performs a local unitary operation $%
V_{B}\left( \mu \right) $ depending on the value of $\mu $, which is defined
by

\begin{equation}
V_{B}\left( \mu \right) =I\cos \theta +i\left( -1\right) ^{\mu }\sigma
_{B}\sin \theta .  \label{vb}
\end{equation}%
Here $\sigma _{B}=\vec{u}_{B}\cdot \vec{\sigma}_{n_{B}}$ , $\vec{u}_{B}$ is
a three-dimensional real unit vector and $\vec{\sigma}_{n_{B}}$ is the Pauli
spin matrix vector at site $n_{B}$. The real parameter $\theta $ is defined
by 
\begin{eqnarray}
\cos \left( 2\theta \right) &=&\frac{\xi }{\sqrt{\xi ^{2}+\eta ^{2}}},
\label{c} \\
\sin (2\theta ) &=&-\frac{\eta }{\sqrt{\xi ^{2}+\eta ^{2}}}.  \label{s}
\end{eqnarray}%
where the real parameters $\xi $ and $\eta $ are given by 
\begin{equation}
\xi =\langle g|\sigma _{B}H\sigma _{B}|g\rangle ,
\end{equation}

\begin{equation}
\eta =\langle g|\sigma _{A}\dot{\sigma}_{B}|g\rangle  \label{12}
\end{equation}%
with

\begin{equation}
\dot{\sigma}_{B}=i\left[ H_{n_{B}},~\sigma _{B}\right] .  \label{dotb}
\end{equation}
For the case with $\eta \neq 0$, Bob obtains positive energy: 
\begin{equation}
E_{B}=\frac{1}{2}\left[ \sqrt{\xi ^{2}+\eta ^{2}}-\xi \right]  \label{enb}
\end{equation}%
on average from the spin chain in the process of the local operation.

\bigskip

The value of $\xi $ is non-negative due to the non-negativity of $H$.
Because of the hermicity and commutativity of $\sigma _{A}$ and $\dot{\sigma}%
_{B} $, the reality of $\eta $ is easily proven as follows:

\begin{eqnarray*}
\eta ^{\ast } &=&\langle g|\sigma _{A}\dot{\sigma}_{B}|g\rangle ^{\ast
}=\langle g|\dot{\sigma}_{B}\sigma _{A}|g\rangle  \\
&=&\langle g|\sigma _{A}\dot{\sigma}_{B}|g\rangle =\eta .
\end{eqnarray*}%
Using $\left[ T_{n},~\sigma _{B}\right] =0$ for $|n-n_{B}|>L$, eq. (\ref%
{dotb}) can be rewritten as%
\begin{equation}
\dot{\sigma}_{B}=i\left[ H,~\sigma _{B}\right] ,  \label{20}
\end{equation}%
and thus $\dot{\sigma}_{B}$ can be interpreted as a time-derivative operator
of the Heisenberg operator $\exp \left[ itH\right] \sigma _{B}\exp [-itH]$
at $t=0$. After Bob's operation, the post-measurement state in eq. (\ref{sa}%
) is transformed into%
\begin{equation*}
\frac{1}{\sqrt{p_{A}(\mu )}}V_{B}\left( \mu \right) P_{A}\left( \mu \right)
|g\rangle .
\end{equation*}%
Here we have neglected time evolution of the spin chain between Alice's
measurement and Bob's operation assuming the relation of eq. (\ref{sts}) or
eq. (\ref{sts2}). The average state is given by

\begin{eqnarray*}
\rho &=&\sum_{\mu =0,1}p_{A}(\mu )\frac{1}{\sqrt{p_{A}(\mu )}}V_{B}\left(
\mu \right) P_{A}\left( \mu \right) |g\rangle \langle g|P_{A}\left( \mu
\right) V_{B}^{\dag }\left( \mu \right) \frac{1}{\sqrt{p_{A}(\mu )}} \\
&=&\sum_{\mu =0,1}V_{B}\left( \mu \right) P_{A}\left( \mu \right) |g\rangle
\langle g|P_{A}\left( \mu \right) V_{B}^{\dag }\left( \mu \right) .
\end{eqnarray*}%
It is straightforward to calculate the average localized energy $\limfunc{Tr}%
\left[ \rho H_{n_{B}}\right] $ after Bob's operation, as follows. Firstly,
by the commutativity of $P_{A}\left( \mu \right) $ and $V_{B}^{\dag }\left(
\mu \right) H_{n_{B}}V_{B}\left( \mu \right) $, we obtain

\begin{eqnarray*}
\limfunc{Tr}\left[ \rho H_{n_{B}}\right]  &=&\sum_{\mu =0,1}\langle
g|P_{A}\left( \mu \right) \left( V_{B}^{\dag }\left( \mu \right)
H_{n_{B}}V_{B}\left( \mu \right) \right) P_{A}\left( \mu \right) |g\rangle 
\\
&=&\sum_{\mu =0,1}\langle g|P_{A}\left( \mu \right) \left( V_{B}^{\dag
}\left( \mu \right) H_{n_{B}}V_{B}\left( \mu \right) \right) |g\rangle ,
\end{eqnarray*}%
where we have used $P_{A}\left( \mu \right) ^{2}=P_{A}\left( \mu \right) $.
Substituting eq. (\ref{vb}) into the above relation yields the following
expression of $\limfunc{Tr}\left[ \rho H_{n_{B}}\right] $: 
\begin{eqnarray*}
&&\limfunc{Tr}\left[ \rho H_{n_{B}}\right]  \\
&=&\sum_{\mu =0,1}\langle g|P_{A}\left( \mu \right) \left( I\cos \theta
-i\left( -1\right) ^{\mu }\sigma _{B}\sin \theta \right) H_{n_{B}}\left(
I\cos \theta +i\left( -1\right) ^{\mu }\sigma _{B}\sin \theta \right)
|g\rangle  \\
&=&\cos ^{2}\theta \langle g|\left( \sum_{\mu =0,1}P_{A}\left( \mu \right)
\right) H_{n_{B}}|g\rangle +\sin ^{2}\theta \langle g|\left( \sum_{\mu
=0,1}P_{A}\left( \mu \right) \right) \sigma _{B}H_{n_{B}}\sigma
_{B}|g\rangle  \\
&&+i\cos \theta \sin \theta \langle g|\left( \sum_{\mu =0,1}\left( -1\right)
^{\mu }P_{A}\left( \mu \right) \right) \left[ H_{n_{B}},~\sigma _{B}\right]
|g\rangle .
\end{eqnarray*}%
By taking account of the completeness relation of $P_{A}\left( \mu \right) $
and the spectral decomposition of $\sigma _{A}$ in eq. (\ref{uas}), we get
the following relation:

\begin{eqnarray*}
\limfunc{Tr}\left[ \rho H_{n_{B}}\right] &=&\cos ^{2}\theta \langle
g|H_{n_{B}}|g\rangle +\sin ^{2}\theta \langle g|\sigma _{B}H_{n_{B}}\sigma
_{B}|g\rangle \\
&&+i\cos \theta \sin \theta \langle g|\sigma _{A}\left[ H_{n_{B}},~\sigma
_{B}\right] |g\rangle .
\end{eqnarray*}%
Because $\langle g|H_{n_{B}}|g\rangle =0$ is satisfied, the form given by 
\begin{equation}
\limfunc{Tr}\left[ \rho H_{n_{B}}\right] =\langle g|\sigma
_{B}H_{n_{B}}\sigma _{B}|g\rangle \sin ^{2}\theta +\eta \cos \theta \sin
\theta  \label{9'}
\end{equation}%
is obtained. Here, the relation:

\begin{eqnarray*}
\langle g|\sigma _{B}H_{n_{B}}\sigma _{B}|g\rangle  &=&\langle g|\sigma
_{B}\left( H-\sum_{n^{\prime }\notin \left[ n_{B}-L,~n_{B}+L\right]
}T_{n^{\prime }}\right) \sigma _{B}|g\rangle  \\
&=&\langle g|\sigma _{B}H\sigma _{B}|g\rangle =\xi 
\end{eqnarray*}%
is satisfied due to the following relation:%
\begin{equation*}
\langle g|\sigma _{B}T_{n^{\prime }}\sigma _{B}|g\rangle =\langle g|\sigma
_{B}\sigma _{B}T_{n^{\prime }}|g\rangle =\langle g|T_{n^{\prime }}|g\rangle
=0
\end{equation*}%
for $n^{\prime }\notin \left[ n_{B}-L,~n_{B}+L\right] $. Thus, eq. (\ref{9'}%
) is rewritten as 
\begin{eqnarray}
\limfunc{Tr}\left[ \rho H_{n_{B}}\right]  &=&\xi \sin ^{2}\theta +\eta \cos
\theta \sin \theta   \notag \\
&=&\frac{\xi }{2}\left( 1-\cos \left( 2\theta \right) \right) +\frac{\eta }{2%
}\sin (2\theta ).  \label{9}
\end{eqnarray}%
The parameter $\theta $ is now fixed to make $\limfunc{Tr}\left[ \rho
H_{n_{B}}\right] $ as negative as possible. This can be achieved by taking $%
\theta $ as in eqs. (\ref{c}) and (\ref{s}). $\limfunc{Tr}\left[ \rho
H_{n_{B}}\right] $ is then evaluated as

\begin{equation}
\limfunc{Tr}\left[ \rho H_{n_{B}}\right] =\frac{1}{2}\left[ \xi -\sqrt{\xi
^{2}+\eta ^{2}}\right]   \label{r1}
\end{equation}%
by substituting eqs. (\ref{c}) and (\ref{s}) into the relation derived from
eq. (\ref{9}). If $\eta \neq 0$, it is clear that $\limfunc{Tr}\left[ \rho
H_{n_{B}}\right] $ in eq. (\ref{r1}) is negative: 
\begin{equation}
\limfunc{Tr}\left[ \rho H_{n_{B}}\right] <0.  \label{10}
\end{equation}%
eq. (\ref{10}) is a significant result. Before step (III), the energy around
Bob is zero:%
\begin{equation*}
\limfunc{Tr}\left[ \rho ^{\prime }H_{n_{B}}\right] =0.
\end{equation*}%
After Bob's local operations in step (III), the localized energy around site 
$n_{B}$ becomes negative. Respecting local energy conservation explained in
section \ 3, this means that positive energy $E_{B}$ must be emitted from
the spin chain to Bob (more precisely, to Bob's devices to perform $%
V_{B}\left( \mu \right) $). More explicitly, local energy conservation
around site $n_{B}$ yields the relation:%
\begin{equation*}
E_{B}+\limfunc{Tr}\left[ \rho H_{n_{B}}\right] =\limfunc{Tr}\left[ \rho
^{\prime }H_{n_{B}}\right] =0.
\end{equation*}%
From this relation, we finally obtain eq. (\ref{enb}) as follows:%
\begin{eqnarray*}
E_{B} &=&\limfunc{Tr}\left[ \rho ^{\prime }H_{n_{B}}\right] -\limfunc{Tr}%
\left[ \rho H_{n_{B}}\right]  \\
&=&\frac{1}{2}\left[ \sqrt{\xi ^{2}+\eta ^{2}}-\xi \right] .
\end{eqnarray*}%
As seen in eq. (\ref{12}), $\eta $ is given by a two-point correlation
function of (semi-)local operators. If the ground state is separable, it
turns out from eq. (\ref{20}) and $H|g\rangle =0$ that $\eta $ vanishes as
follows:%
\begin{eqnarray*}
\eta  &=&\langle g|\sigma _{A}\dot{\sigma}_{B}|g\rangle =\langle g|\sigma
_{A}|g\rangle \langle g|\dot{\sigma}_{B}|g\rangle  \\
&=&i\langle g|\sigma _{A}|g\rangle \langle g|\left( H\sigma _{B}-\sigma
_{B}H\right) |g\rangle =0.
\end{eqnarray*}%
Therefore, it can easily be checked from eq. (\ref{enb}) that Bob gains no
energy, as should be the case. However, the correlation function $\eta $
does not vanish in general because entanglement yields correlations among
the quantum fluctuations at each point. For example, the critical Ising spin
chains with transversal magnetic field have nonvanishing $\eta $ \cite%
{hotta2}. In general, any spin chain models without fine-tuning may have
nonvanishing $\eta $, just like the Ising spin chains, and the protocol in
this paper becomes effective barring fine-tuned exceptions. Schematic plots
of the expectation values of energy density for the steps of the QET are
given in Figs. 2--4. From these figures, it can be easily understood that
there exists no energy flow between Alice and Bob. Energy is locally
transported from the spin chain to Bob's device for the operation by
simultaneously generating negative energy $-E_{B}$ in the spin chain and
positive energy $+E_{B}$ in the device of Bob. Hence, local energy
conservation is exactly maintained in the protocol.

It is worth commenting here that the ground state $|g\rangle $ has a typical
correlation site length $l$, over which the correlation between two sites
and the value of $\eta ~$decay rapidly. Hence, the protocol is more
effective in teleporting energy between Alice and Bob with $\left\vert
n_{A}-n_{B}\right\vert \lessapprox l$, because the energy gain of Bob
increases when $\eta $ is large. To achieve long-range QET, it is preferable
to choose near-critical spin chain systems which have entangled ground
states with long-range correlation, that is, $l\gg 1$. For example, the
near-critical Ising spin chains in the presence of transversal magnetic
field with $b\sim h>0$ in eq. (\ref{ising}) may be good candidates.

In the above analysis, we have shown that Bob obtains energy from the spin
chain. However, even after the last step of the protocol, there exists
energy $E_{A}$ that Alice had to first deposit to the spin chain by her
measurement. Actually, using $H_{n_{A}}V_{B}\left( \mu \right) =V_{B}\left(
\mu \right) H_{n_{A}}$, we are able to check easily that $\limfunc{Tr}\left[
\rho H_{n_{A}}\right] =\limfunc{Tr}\left[ \rho ^{\prime }H_{n_{A}}\right] $.
Let us imagine that Alice attempts to completely withdraw $E_{A}~$by local
operations after step (III). If this was possible, the energy gain $E_{B}$
of Bob might have no cost. However, if so, the total energy of the spin
chain becomes equal to $-E_{B}$ and negative. Meanwhile, we know that the
total energy of the spin chains must be nonnegative. Hence, Alice cannot
withdraw energy larger than $E_{A}-E_{B}$ by local operations at site $n_{A}$%
. The reason of existence of the residual energy has been already explained
in section 4 from the viewpoint of entanglement breaking. It may be
instructive to stress the simple fact that if the deposited energy $E_{A}$
vanishes, $E_{B}$ also vanishes. \ This implies that Bob releases a part of
energy poured by Alice's measurement.

\section{Conclusion}

\bigskip

In this paper, a protocol of QET is proposed in spin chain systems that
respects all physical laws including causality and local energy
conservation. Energy is effectively transported from Alice to Bob by only
LOCC, using local excitations with negative energy and ground-state
entanglement. The energy input by Alice is given by eq. (\ref{ea}) and the
energy gain by Bob is given by eq. (\ref{enb}).

If we do not consider the ground state, but rather an excited state of the
spin chain which has nonzero energy distribution around Bob, it is not
anomalous that Bob can extract energy from the spin chain. In the QET
protocol, the initial \ state is set to be the ground state and energy
extraction by Bob is nontrivial. At first glance, energy extraction from the
ground state by Bob appears impossible due to local energy conservation
because Alice only excites locally the spin chain by her measurements. Based
on this, we may wonder why energy is teleported in QET without any physical
agent in the protocol. The answer may be briefly summarized as follows. As
stressed in section 2, quantum mechanics allows negative-energy-density
regions by controlling quantum fluctuation. Hence, even if Bob has no energy
on average around him, the value of the energy density around him can
decrease. Therefore, the ground state can be locally regarded as
\textquotedblleft an excited state\textquotedblright , compared with the
state with negative energy around Bob. In the QET protocol, Bob extracts
this local excitation energy by using Alice's measurement result. In this
way, it can be said that the energy, which Bob will obtain, existed around
Bob \textit{before} the start of the QET protocol. Therefore, we do not need
any transfer of energy from Alice to Bob. Of course, the energy hidden
behind Bob's region is not always available. In the protocol, classical
information about the measurement result of Alice becomes a key to extract
energy by Bob. Without knowledge about the measurement result, Bob cannot
get any energy from the local ground state. The essential reason why the
measurement result allows Bob to get energy is as follows. Because the spin
of Alice is entangled with spins around Bob, the measurement result includes
information about the quantum fluctuations around Bob. Bob infers from the
measurement information how the quantum fluctuations behave around him and
can choose a unitary operation to extract but not give energy from the spin
chain. Consequently, the protocol is able to transport energy by LOCC
without breaking local energy conservation and without the existence of any
excited physical entity.

\bigskip

\bigskip

\bigskip

\textbf{Acknowledgments}\newline

\bigskip

This research was partially supported by the SCOPE project of the MIC.

\bigskip

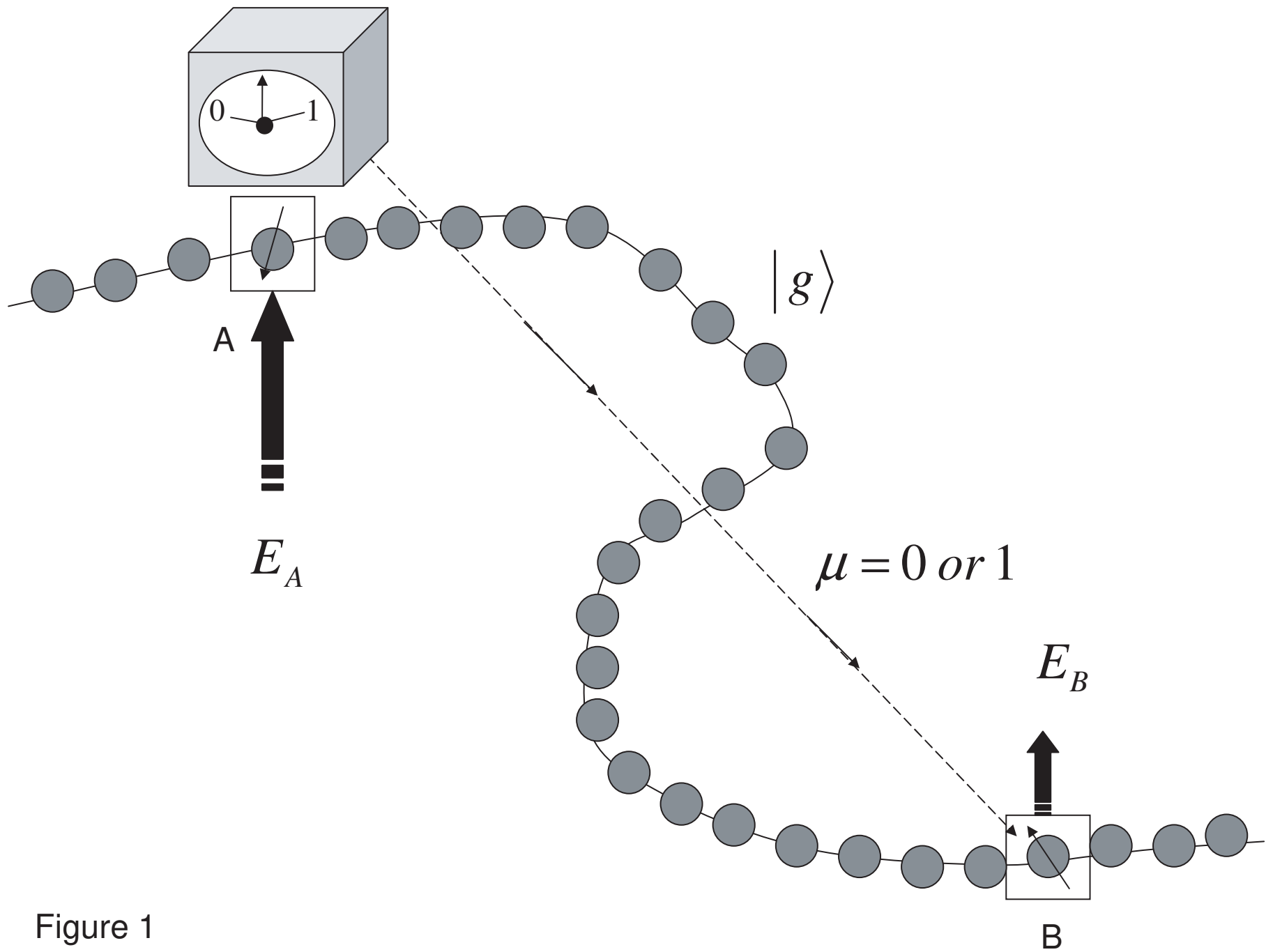
Figure 1: QET protocol

\bigskip

(1) The eigenvalues of $\vec{u}_{A}\cdot \vec{\sigma}_{A}$ are $\left(
-1\right) ^{\mu }$ with $\mu =0,1$. Alice performs a local projective
measurement of the observable $\vec{u}_{A}\cdot \vec{\sigma}_{A}$ to her
spin in the ground state $|g\rangle $ and obtains the measurement result $%
\mu $. Alice must input energy $E_{A}$ to the spin chain system in order to
achieve the local measurement. (2) Alice announces to Bob the result $\mu $
by a classical channel. (3) Bob performs a local unitary operation $%
V_{B}\left( \mu \right) $ to his spin, depending on the value of $\mu $.~Bob
obtains energy output $E_{B}~$on average from the spin chain system in the
process of the local operation.

\bigskip

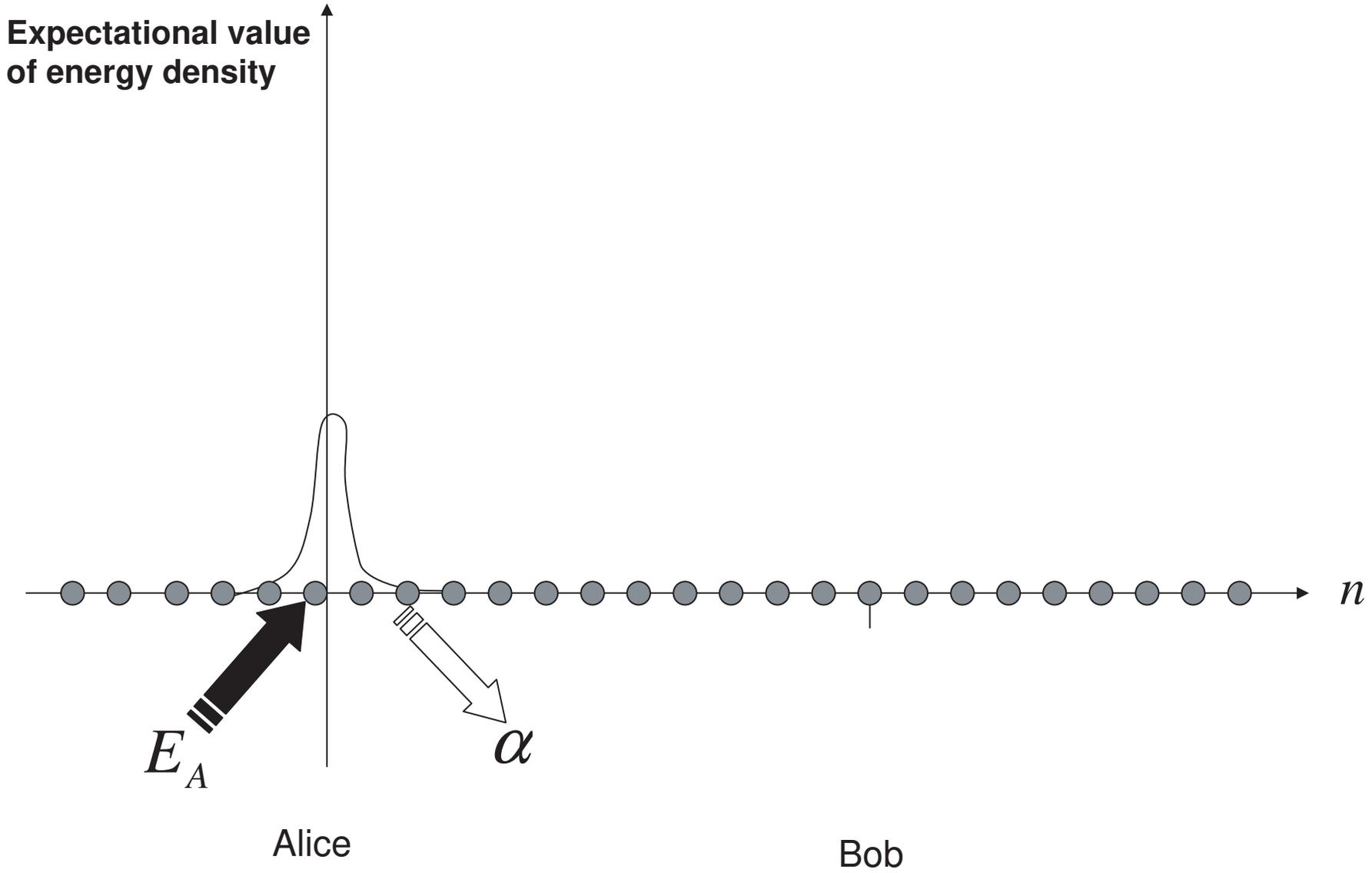
Figure 2: Schematic plot of the expectation value of energy density in Step
(I). Alice measures $\vec{u}_{A}\cdot \vec{\sigma}_{A}$ and obtains the
result $\mu $. The energy $E_{A}$ is poured into the spin chain by the local
measurement.

\bigskip

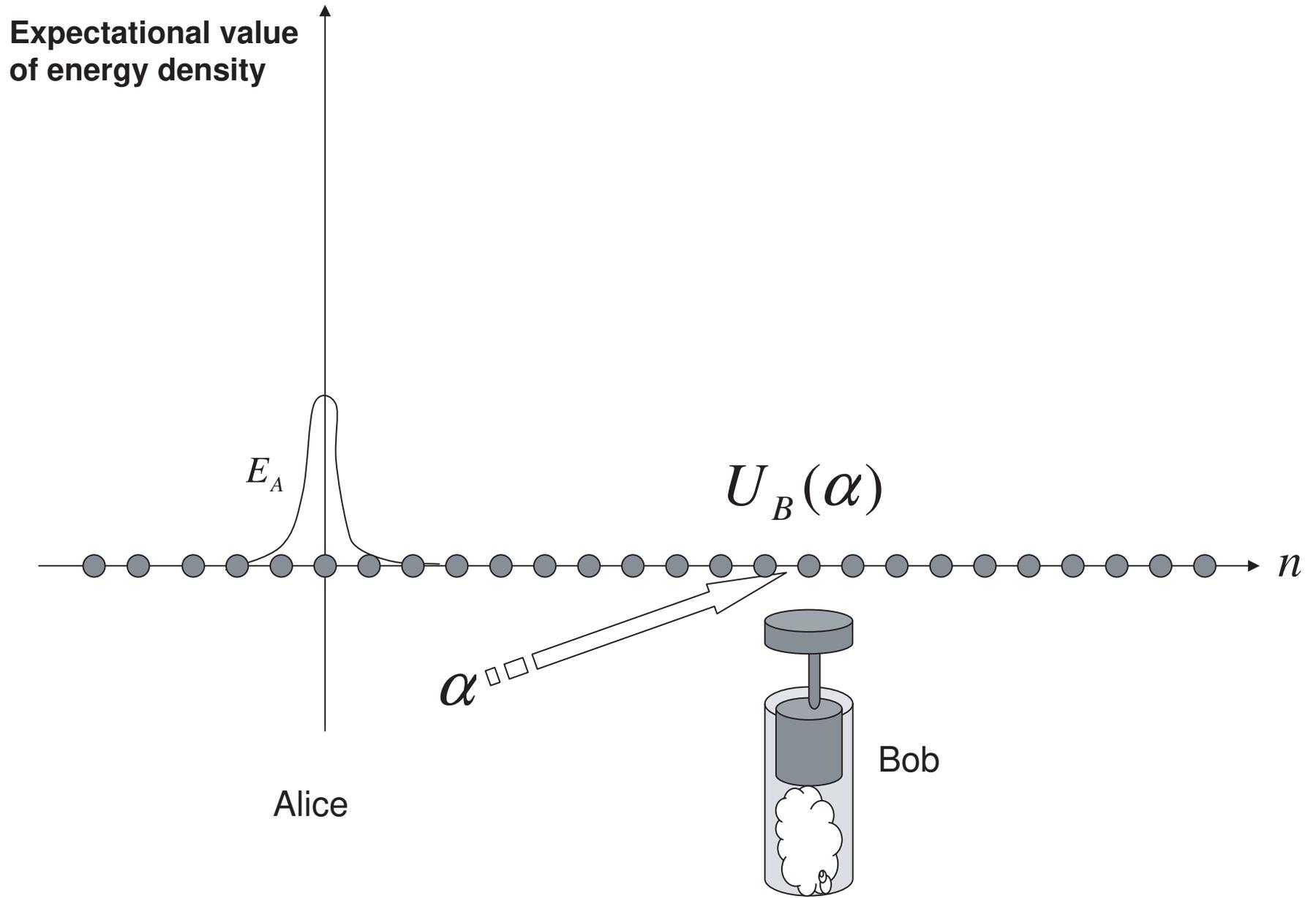
Figure 3: Schematic plot of the expectation value of energy density in Step
(II). Alice announces $\mu $ to Bob. Bob perfomes result-dependent unitary
operation to his spin.

\bigskip

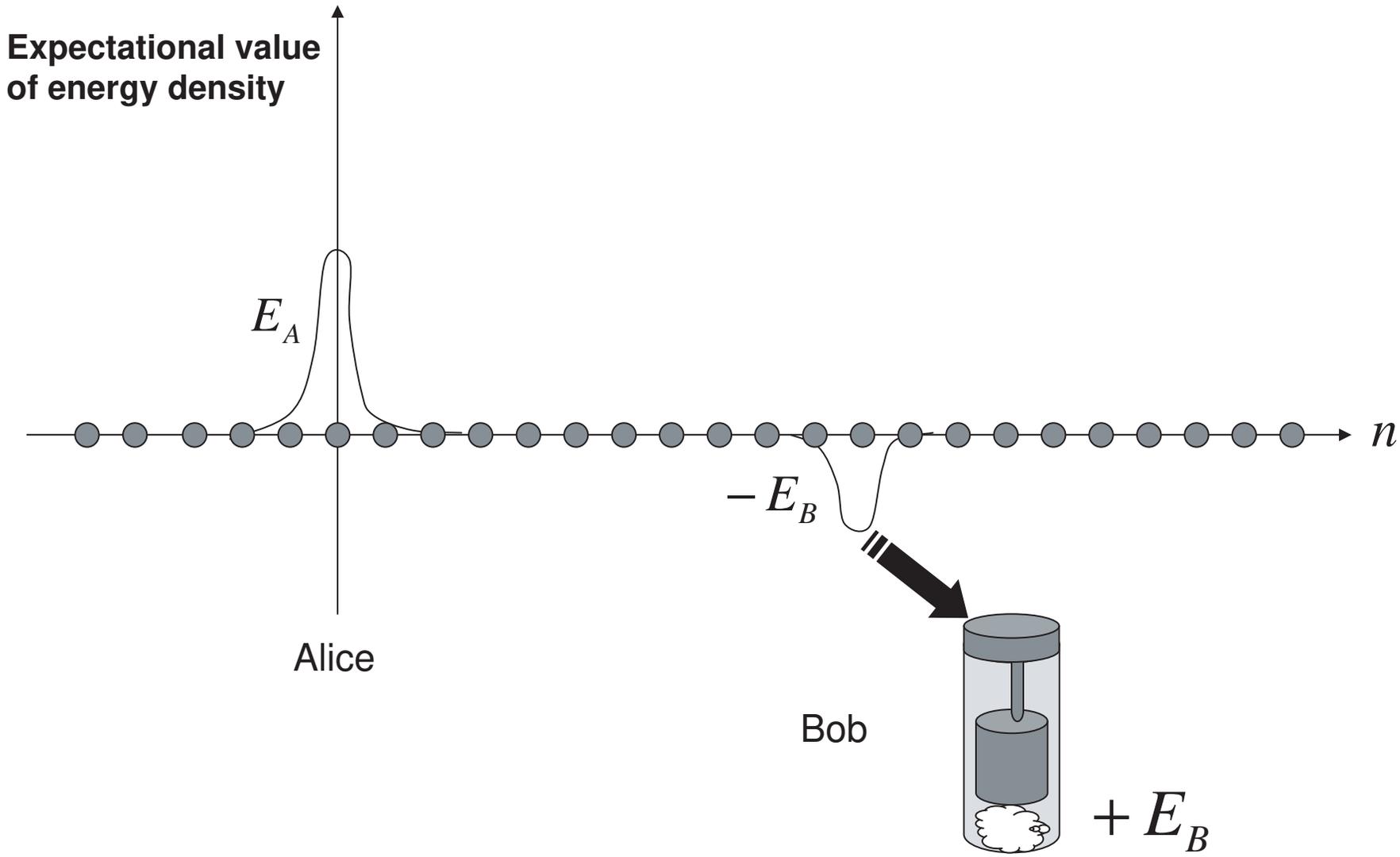
Figure 4: Schematic plot of the expectation value of energy density in Step
(III). Negative energy $-E_{B}$ appears in the spin chain and positive
energy $+E_{B}$ is released outside. Bob is able to use the energy.

\bigskip


\bigskip

\begin{thebibliography}{9}
\bibitem{qt} C.H. Bennett, G. Brassard, C. Cr\'{e}peau, R. Jozsa, A. Peres,
and W.K. Wootters: Phys. Rev. Lett. 70 (1993) 1895.

\bibitem{stqi} S. Bose: Phys. Rev. Lett. 91 (2003) 207901; F. Verstraete,
M.A. Mart`\i n-Delgado, and J.I. Cirac: Phys. Rev. Lett. 92 (2004) 087201.

\bibitem{rev} L. Amico, R. Fazio, A. Osterloh, and V. Vedral:
quant-ph/0703044.

\bibitem{BD} N.D. Birrell and P.C.W. Davies: \textit{Quantum Fields in
Curved Space} (Cambridge Univ. Press, 1982) p. 269.

\bibitem{hotta} M. Hotta: Phys. Lett. A372 (2008) 3752.

\bibitem{hotta2} M. Hotta: Phys. Lett. A372(2008) 5671.

\bibitem{P} P. Pfeuty: Ann. Phys. 57(1970)79.

\newpage

Figure Caption
\end{thebibliography}
\end{document}